# Non-dipole effects in a multielectron atom

Alexey N. Hopersky, Alexey M. Nadolinsky, and Rustam V. Koneev

Rostov State Transport University, 344038, Rostov-on-Don, Russia

*E-mail*: qedhop@mail.ru, amnrnd@mail.ru, koneev@gmail.com

---

**Abstract.** This Letter establishes the analytical structure of the dipole radiation transition operator, taking into account non-dipole effects for a multielectron atom. The results summarize those of Komninos Y. *et al.*, Phys. Rev. A **86**, 023420 (2012), obtained in an approximation of a "split" electron of a continuous spectrum. The result for the complete operator of a multipole radiation transition [see formula (10)] can be specified within the framework of the methods of the theory of irreducible tensor operators and for higher multipoles.

---

Let us consider the *integral representation of Loudon* [1] for the electrical part of the complete operator of the interaction of an electromagnetic field with a multielectron atom (non-relativistic approximation):

$$H_e = e\sum_{n=1}^{N}\int_0^1 d\lambda(\vec{r}_n\cdot\vec{E}(\lambda\vec{r}_n\cdot\vec{k})). \quad (1)$$

In (1) $e$ – electron charge, $N$ is the number of electrons in the atom, $\vec{r}_n$ is the radius-vector of the *n*-electron of the atom, $\vec{k}$ is the wave vector of the photon, and the electric field operator (at time $t$ = 0) is defined (in the secondary quantization representation):

$$\vec{E}(\vec{r}_n,0) = i(1/c)\sum\nolimits_{\vec{k}\rho}\omega_k\vec{e}_{\vec{k}\rho}(a^+_{\vec{k}\rho}e^{-i(\vec{k}\cdot\vec{r}_n)} - a^-_{\vec{k}\rho}e^{i(\vec{k}\cdot\vec{r}_n)}). \quad (2)$$

B (2) $c$ – the speed of light in a vacuum, $\rho$ – polarization index, $\vec{e}_{\vec{k}\rho}(\omega_k)$ – polarization vector (energy), $a^+_{\vec{k}\rho}$ $(a^-_{\vec{k}\rho})$ – photon production (annihilation) operator, and $(\vec{e}_{\vec{k}\rho}\cdot\vec{k}) = 0$. Directing the photon wave vector along the axis $OZ$, we have the *Rayleigh decomposition* for the exponent:

$$e^{\pm i(\vec{k}\cdot\vec{r}_n)} = \sum_{l=0}^{\infty}(\pm i)^l j_l(kr_n)\sqrt{4\pi(2l+1)}\cdot Y_{l0}(\vec{e}_n), \quad (3)$$

where $j_l$ – spherical Bessel function, $Y_{l0}$ – spherical harmonica, $k = \vec{k}/|\vec{k}|$, $\vec{e}_n = \vec{r}_n/r_n$, $r_n = |\vec{r}_n|$ and $Y^*_{l0} = Y_{l0}$. Consider the representation of the dot product in terms of spherical functions:

$$(\vec{e}_{\vec{k}\rho}\cdot\vec{r}_n) = r_n\sum_p(-1)^p C^{(1)}_{-p}(\vec{e}_{\vec{k}\rho})C^{(1)}_p(\vec{e}_n), \quad p = -1, 0, 1, \quad (4)$$

$$C^{(1)}_p(\vec{e}_n) = \sqrt{4\pi/3}\cdot Y_{1p}(\vec{e}_n). \quad (5)$$

Then for operator (1) we get:

$$H_e = i\frac{e}{c}(4\pi/3)\sum_{n=1}^{N}\sum_{\vec{k}\rho}\sum_p\sum_{l=0}^{\infty}\omega_k(-1)^p C^{(1)}_{-p}(\vec{e}_{\vec{k}\rho})\cdot i^l\cdot r_n\cdot$$
$$\cdot\sqrt{2l+1}\cdot[(-1)^l a^+_{\vec{k}\rho} - a^-_{\vec{k}\rho}]\cdot\int_0^1 d\lambda\cdot j_l(\lambda kr_n)Y_{1p}(\vec{e}_n)Y_{l0}(\vec{e}_n). \quad (6)$$

The product of spherical harmonics in (6) is represented by the Klebsch-Gordan expansion:

$$Y_{1p}(\vec{e}_n)Y_{l0}(\vec{e}_n) = \sum_{L=0}^{\infty}\sum_{m=-L}^{L}\sqrt{\frac{3(2l+1)}{4\pi(2L+1)}}\cdot C^{L0}_{10l0}C^{LM}_{1pl0}\cdot Y_{LM}(\vec{e}_n), \quad (7)$$

where the Klebsch–Gordan coefficients are defined by $3j$–Wigner symbols:

$$C^{L0}_{10l0} = (-1)^{l+1}\sqrt{2L+1}\cdot\begin{pmatrix}1 & l & L\\ 0 & 0 & 0\end{pmatrix}, \tag{8}$$

$$C^{LM}_{1pl0} = (-1)^{l+M+1}\sqrt{2L+1}\cdot\begin{pmatrix}1 & l & L\\ p & 0 & -M\end{pmatrix}, \tag{9}$$

Substituting (7) and communication $Y_{LM}(\vec{e}_n) = \sqrt{(2L+1)/4\pi}\cdot C^{(L)}_M(\vec{e}_n)$ in (6), for a complete operator of a multipole radiation transition in a multielectron atom, we have:

$$\begin{aligned} H_e = i\frac{e}{c}\sum_{\vec{k}\rho}\sum_{p}\sum_{l=0}^{\infty}\sum_{LM}\omega_k(-1)^{p+M}C^{(1)}_{-p}(\vec{e}_{\vec{k}\rho})\cdot i^l\cdot(2l+1)(2L+1)\cdot \\ \cdot[(-1)^l a^+_{\vec{k}\rho} - a^-_{\vec{k}\rho}]\cdot\begin{pmatrix}1 & l & L\\ 0 & 0 & 0\end{pmatrix}\begin{pmatrix}1 & l & L\\ p & 0 & -M\end{pmatrix}\cdot\int_0^1 d\lambda\cdot\sum_{n=1}^{N} r_n j_l(\lambda k r_n) C^{(L)}_M(\vec{e}_n). \end{aligned} \tag{10}$$

Note that for $L = 0$ ($M = 0$) by virtue of the condition on the bottom line $3j$-Wigner symbol ($p - M = 0$) we have only $p = 0$. Then, with $\vec{k}\parallel OZ$ component with $L = 0$ ($l = 1$) in (10) disappears.

An analytical summation of the series in (10) will be carried out only for the dipole term: $L = 1$, $l = 0, 2$ (see the selection rules for the top row $3j$-Wigner symbol: $L \geq |l-1|$, $L \leq l+1$, $1+l+L$ – even). We get:

$$\begin{aligned} H^{(D)}_e = 3i\cdot\frac{e}{c}\cdot\sum_{\vec{k}\rho}\sum_{p}\sum_{M}\omega_k(-1)^{p+M}C^{(1)}_{-p}(\vec{e}_{\vec{k}\rho})(a^+_{\vec{k}\rho} - a^-_{\vec{k}\rho})\cdot \\ \cdot\sum_{n=1}^{N} r_n\cdot C^{(1)}_M(\vec{e}_n)\cdot\int_0^1 d\lambda\cdot\left\{\begin{pmatrix}1 & 0 & 1\\ 0 & 0 & 0\end{pmatrix}\begin{pmatrix}1 & 0 & 1\\ p & 0 & -M\end{pmatrix}j_0(\lambda k r_n) - \right. \\ \left. -5\cdot\begin{pmatrix}1 & 2 & 1\\ 0 & 0 & 0\end{pmatrix}\begin{pmatrix}1 & 2 & 1\\ p & 0 & -M\end{pmatrix}j_2(\lambda k r_n)\right\}. \end{aligned} \tag{11}$$

Considering the values of (a) $3j$-Wigner's symbols ($\delta_{pM}$ – Kronecker symbol)

$$\begin{pmatrix}1 & 0 & 1\\ 0 & 0 & 0\end{pmatrix} = -\frac{1}{\sqrt{3}},\quad \begin{pmatrix}1 & 0 & 1\\ p & 0 & -M\end{pmatrix} = (-1)^{M+1}\cdot\frac{1}{\sqrt{3}}\cdot\delta_{pM}, \tag{12}$$

$$\begin{pmatrix}1 & 2 & 1\\ 0 & 0 & 0\end{pmatrix} = \frac{2}{\sqrt{30}},\quad \begin{pmatrix}1 & 2 & 1\\ p & 0 & -M\end{pmatrix} = (-1)^{p+1}\cdot\frac{(3p^2-2)}{\sqrt{30}}\cdot\delta_{pM}, \tag{13}$$

(b) photonic spherical functions for the fixed scheme of the intended experiment ($\vec{k}\parallel OZ$, $\vec{e}_{\vec{k}\rho}\perp OZ$) on the interaction of X-rays with a multielectron atom

$$C^{(1)}_0(\vec{e}_{\vec{k}\rho}) = 0,\ C^{(1)}_{\pm1}(\vec{e}_{\vec{k}\rho}) = \mp\sqrt{3/8\pi}, \tag{14}$$

(c) recurrence relation for spherical Bessel functions

$$j_0(z) + j_2(z) = (3/z)j_1(z), \tag{15}$$

and summarizing by $p$– and $M$–parameters, for the dipole junction operator (11) we finally get:

$$H^{(D)}_e = i\frac{e}{c}\cdot\sqrt{3/8\pi}\cdot\sum_{\vec{k}\rho}\omega_k\cdot(a^+_{\vec{k}\rho} - a^-_{\vec{k}\rho})\cdot Q, \tag{16}$$

$$Q = \sum_{n=1}^{N}[C^{(1)}_{-1}(\vec{e}_n) - C^{(1)}_1(\vec{e}_n)]\cdot\Psi_n, \tag{17}$$

$$\Psi_n = (3/k)\int_0^{r_n} \frac{dx_n}{x_n} \cdot j_1(kx_n), \quad x_n = \lambda r_n. \tag{18}$$

Formula (16) gives an accurate result for the operator $H_e^{(D)}$. Let's find a simple approximation for it. With $kr_n << 1$ we have a dipole approximation regime:

$$\Psi_n \to (1/k) \cdot Si(kr_n) \to r_n, \tag{19}$$

$$Si(kr_n) = \int_0^{kr_n} dz \cdot j_0(z). \tag{20}$$

With $kr_n >> 1$ we have asymptotics of the integral sinus function:

$$\Psi_n \to (3/2k) \cdot Si(kr_n), \tag{21}$$

$$Si(kr_n) \to \pi/2. \tag{22}$$

As a result, the operator $Q$ in (16) takes on a simple form:

$$Q \cong \sum_{n=1}^{N} [C_{-1}^{(1)}(\vec{e}_n) - C_1^{(1)}(\vec{e}_n)] \cdot D_n, \tag{23}$$

$$D_n = \begin{cases} r_n, \; r_n \in [0; q), \\ q, \; r_n \in [q; \infty), \end{cases} \tag{24}$$

$$q = \frac{3}{2k}\frac{\pi}{2} = (3/8)\lambda_\omega, \quad k = 2\pi/\lambda_\omega, \tag{25}$$

where $\lambda_\omega$ – the wavelength of the photon. According to (24), parameter $q$ plays the role of a "breakpoint" for the infinite growth of the operator $r_n$.

Formulas (18) and (24) give a generalization of formulas (6) and (9) of work [2] in the case of a multielectron atom. As an example of the use of approximation (24), we note the discovery of a *giant non-dipole effect* in the theoretical description of generalized cross-sections of two-photon double ionization of the $K$–shell of a neon-like atomic ion ($Fe^{16+}$) [3]. For the amplitude of the probability of a radiation transition between single-electron states of a continuous spectrum, the dipole approximation loses its physical meaning at an infinite interval $r_n \in [0; \infty)$ due to the non-localizability of the continuous spectrum. In this case, you must use the operator $Q$ from (17) or from (23). The study of the analytical structure of the terms of the complete operator (10) for higher multipoles ($L \geq 2$) by the methods of the theory of irreducible tensor operators is the subject of the future development of the theory of non-dipole effects in a multielectron atom. At the same time, each value of $L$ from the counting set [1; ∞) corresponds to only two values of $l = L-1,\; L+1$ (for example, for the quadrupole transition operator $L = 2$ and $l = 1, 3$) due to the conditions on the strings of the 3*j*–Wigner symbol. In other words, formally mathematically infinite series on $L \in [1; \infty)$ and $l \in [0; \infty)$ in (10) turn out to be finite sums.

--------------------------------------------------------------------------------------------------------